\documentclass[prb,preprint,showpacs,preprintnumbers,amsmath,amssymb]{revtex4}

\usepackage[dvipdfmx]{graphicx}
\usepackage{bm}% bold math

\newcommand{\be}{\begin{equation}}
\newcommand{\ee}{\end{equation}}
\newcommand{\eq}[1]{Eq.~(\ref{#1})}
\newcommand{\fig}[1]{Fig.~\ref{#1}}
\def\bea{\begin{eqnarray}}
\def\eea{\end{eqnarray}}

\def\vq{{\bf q}}

\def\vk{{\bf k}}

\begin{document}

\title{Doping dependence of  {\boldmath $d$}-wave bond-charge 
excitations in electron-doped cuprates} 

 \author{Hiroyuki Yamase$^\ddag$, Mat\'{\i}as Bejas$^\dag$, and Andr\'es Greco$^\dag$}
 \affiliation{
  {$^\ddag$}National Institute for Materials Science, Tsukuba 305-0047, Japan \\
 {$^\dag$}Facultad de Ciencias Exactas, Ingenier\'{\i}a y Agrimensura and
 Instituto de F\'{\i}sica Rosario (UNR-CONICET),
 Av. Pellegrini 250, 2000 Rosario, Argentina\\		
 }

\date{\today}

\begin{abstract}
Motivated by the recent experiments reporting the doping dependence of the short-range 
charge order (CO) in electron-doped cuprates, we study the resonant x-ray scattering spectrum 
from $d$-wave bond-charge fluctuations obtained in the two-dimensional $t$-$J$ model. 
We find that (i) the CO is short-range, 
(ii) the CO peak is pronounced at low temperature, 
(iii) the peak intensity increases with decreasing carrier doping $\delta$ down to $\delta \approx 0.10$ 
and is substantially suppressed below $\delta \approx 0.10$ due to strong damping, and 
(iv) the momentum of the CO decreases 
monotonically down to $\delta \approx 0.10$ and goes up below $\delta \approx 0.10$. 
These results reasonably capture the major features of the experimental data, and 
the observed short-range CO can be consistently explained in terms of 
bond-charge fluctuations with an internal $d$-wave symmetry. 
\end{abstract}

\maketitle
\section{introduction}
Recently resonant x-ray scattering (RXS), resonant inelastic 
x-ray scattering (RIXS), and high-energy x-ray scattering revealed 
the presence of short-range charge order (CO) with modulation vector along the axial direction 
$(0,0)$-$(\pi,0)$  in various hole-doped cuprates (h-cuprates)
such as Y- \cite{ghiringhelli12, chang12, achkar12, blackburn13, blanco-canosa14}, 
Bi- \cite{comin14,da-silva-neto14,hashimoto14,peng16}, and Hg-based \cite{gerber15,tabis17} compounds, 
implying that the CO can be a universal phenomenon in h-cuprates.
The understanding of the origin of those charge correlations, therefore, will likely 
yield an important clue to the origin of the pseudogap as well 
as high-$T_c$ superconductivity\cite{keimer15}. 
In fact, a large number of theoretical studies were performed \cite{bejas12,sachdev13,allais14,efetov13,meier14,
wang14,mishra15,yamakawa15,atkinson15,atkinson16}, although a consensus has not been obtained. 

On the other hand, a short-range CO was also observed in electron-doped cuprates
\cite{da-silva-neto15,da-silva-neto16,da-silva-neto18} (e-cuprates).  
Since the pseudogap features similar to those in h-cuprates are much weaker 
in e-cuprates, a theoretical study may be less complicated in e-cuprates. 
However, compared to theoretical studies of h-cuprates \cite{bejas12,sachdev13,allais14,efetov13,meier14,
wang14,mishra15,yamakawa15,atkinson15,atkinson16}, 
the  CO in e-cuprates is much less studied \cite{bejas14,yamase15b,li17,mou17}. 
Ref.~\onlinecite{bejas14} showed a comprehensive study of all possible COs in the 
two-dimensional (2D) $t$-$J$ model and found a strong tendency to $d$-wave bond-charge 
order \cite{misc-d-wave-bond}. 
Ref.~\onlinecite{yamase15b}  then showed that $d$-wave bond-charge fluctuations can 
capture the charge excitation spectrum observed in experiments \cite{da-silva-neto15}. 
Although the theoretical framework is different from Refs.~\onlinecite{bejas14} and \onlinecite{yamase15b}, 
similar $d$-wave bond-charge fluctuations were also proposed in Ref.~\onlinecite{li17} to explain the 
experimental data.

The $d$-wave bond-charge order is different from a usual textbook-like charge-density-wave  
because the bond charge has an internal structure 
characterized by a $d$-wave symmetry. 
Therefore if the short-range CO observed in 
e-cuprates is indeed a $d$-wave bond-charge order, it 
can be interpreted as the first observation of unconventional CO in e-cuprates. 
Given that $d$-wave bond-charge order was discussed in h-cuprates \cite{comin15}, 
it can be a universal phenomenon in the whole cuprate family. 
In addition, $d$-wave bond-charge order would be reduced to the electronic nematic order \cite{kivelson98}, 
more precisely a $d$-wave Pomeranchuk instability \cite{yamase00a,yamase00b,metzner00} 
when the momentum transfer approaches zero. In this sense, the nematic physics can also play a role 
in the charge dynamics in e-cuprates while the nematic physics is discussed 
only in h-cuprates so far \cite{hinkov08,daou10,cyr-choiniere15,sato17}. 
Therefore it is very important to study whether recent experimental data in e-cuprates \cite{da-silva-neto16}, 
i.e., the doping dependence of the short-range CO, can be indeed 
captured in terms of $d$-wave bond charge, which will then  
provide a crucial step to establish the presence of $d$-wave bond-charge fluctuations in e-cuprates. 

In this paper,  we compute the static charge susceptibility associated with $d$-wave 
bond-charge order and then the equal-time correlation function $S(\vq)$, 
the quantity measured by RXS, in a large-$N$ scheme of the $t$-$J$ model. 
We show that our theoretical results capture the major features of the doping dependence 
of the recent RXS data \cite{da-silva-neto16}. 
Our calculations suggest the presence of unconventional charge fluctuations 
in e-cuprates, which are connected with the nematic order 
in the limit of momentum to zero.

\section{model and formalism}
The $d$-wave bond charge is obtained in a non-bias study of the 2D 
$t$-$J$ model by employing a large-$N$ scheme \cite{bejas12}. We follow 
such a theoretical scheme and focus on the excitation spectrum of the $d$-wave bond charge. 

The 2D $t$-$J$ model on a square lattice reads  
\begin{widetext}
\begin{equation}
H = -\sum_{i, j,\sigma} t_{i j}\tilde{c}^\dag_{i\sigma}\tilde{c}_{j\sigma} + 
J \sum_{\langle i,j \rangle}  \left( \vec{S}_i \cdot \vec{S}_j - \frac{1}{4} n_i n_j \right)
+ V \sum_{\langle i,j \rangle} n_i n_j \, 
\label{tJV}  
\end{equation}
\end{widetext}
where $t_{i j}=t(t')$ is the hopping between the first (second) nearest-neighbors sites; 
$J$ and $V$ are the magnetic exchange and Coulomb interaction between the 
nearest-neighbors sites as denoted by $\langle i,j \rangle$, respectively. 
$\tilde{c}^\dag_{i\sigma}$ ($\tilde{c}_{i\sigma}$) is 
the creation (annihilation) operators of electrons with spin $\sigma(=\uparrow, \downarrow)$  
in the Fock space without any double occupancy. 
$n_i=\sum_{\sigma} \tilde{c}^\dag_{i\sigma}\tilde{c}_{i\sigma}$ 
is the electron density operator and $\vec{S}_i$ is the spin operator.

Various approximations to the $t$-$J$ model \cite{gooding94,martins01,bejas14} found 
a strong tendency toward phase separation especially for band parameters appropriate for 
e-cuprates. The phase separation, however, can be an artifact caused by 
discarding the Coulomb repulsion\cite{greco17}. Therefore we included the nearest-neighbor Coulomb 
interaction in the Hamiltonian (\ref{tJV}) to suppress the tendency of the phase separation. 
From a more realistic point of view, we would include the interlayer hopping as well as 
the long-range Coulomb interaction. This is actually important when studying the high-energy 
charge excitation spectrum, which is dominated by plasmon excitations \cite{greco16,bejas17,greco18}. 
However, three dimensionality and the precise form of the Coulomb interaction are not important 
to low-energy charge  excitations  \cite{bejas17}  
addressed in the present work. 

It is not straightforward to analyze the Hamiltonian (\ref{tJV}) because it is defined in the Fock 
space without double occupancy of electrons. 
Here we employ a large-$N$ technique in a path integral representation of the 
Hubbard $X$ operators \cite{foussats04,bejas12}, where the leading order approximation 
becomes exact in the limit of large $N$. With this formalism, all possible charge instabilities 
included in the Hamiltonian (\ref{tJV}) are treated on equal footing and were 
studied at leading order \cite{bejas12, bejas14}. 
In particular, for band parameters appropriate for e-cuprates, $d$-wave bond-charge instability 
is leading around the doping rate $\delta=0.15$; 
see Figs.~4(b) and 6 in Ref.~\onlinecite{bejas14} for the phase diagram. 
As seen in the phase diagram, 
there are other CO tendencies close to $d$-wave bond-charge order 
and their low-energy charge excitations are actually present; see Fig.~7 in Ref.~\onlinecite{bejas17}. 
However, other charge excitations do not show a possible softening along the $(0, 0)$-$(\pi, 0)$ direction and 
cannot capture the experimental data \cite{da-silva-neto15}.  In fact, it is only $d$-wave bond charge 
which exhibits a peak structure along the $(0, 0)$-$(\pi, 0)$ direction in $S(\vq)$ as shown in Fig.~6 in Ref.\onlinecite{bejas17} and captures \cite{yamase15b} the essential features of the experimental 
data in e-cuprates at $\delta=0.14$ and $0.15$ (Ref.~\onlinecite{da-silva-neto15}).  
Therefore to study the recent experimental data performed at 
different doping rates in e-cuprates  \cite{da-silva-neto16}, we focus on the charge excitations 
coming from $d$-wave bond-charge fluctuations. 
Since we deal with the $t$-$J$ model, which is derived from the three-band Hubbard model\cite{fczhang88},
bond-charge order can be interpreted as charge-density-wave at the oxygen sites 
because bond charge is defined on a bond between the nearest-neighbor Cu sites. 

RXS measures the equal-time correlation function, which is defined by 
\be
S(\vq)=\frac{1}{\pi} \int_{-\omega_c}^{\omega_c}  {\rm d}\omega \, {\rm Im} \chi_{d}(\vq,\omega) 
\left[ n_B(\omega)+1 \right], 
\label{Sq}
\ee
where $n_B(\omega)=1/({\rm e}^{\omega/T}-1)$ is the Bose factor and $T$ is temperature. 
We introduced the cutoff energy $\omega_c$ for a later convenience (see \fig{Sq_cut}) 
and $\omega_c=\infty$ for the equal-time correlation function. 
The $d$-wave bond-charge susceptibility $\chi_d(\vq,\omega)$ is obtained in the large-$N$ 
expansion at leading order as \cite{bejas12} 
\be
\chi_{d}(\vq, \omega)= \frac{(8J\Delta^2)^{-1}} {1-2J\Pi_{d}(\vq, \omega)} \,, 
\label{chid}
\ee
which becomes exact in the limit of large $N$. Here 
$\Delta$ is the mean-field value of a bond field and is given by 
$\Delta= \frac{1}{4N_s} \sum_{\vk} (\cos k_x + \cos k_y) f(\epsilon_{\vk})$; 
the value of $\Delta$ is determined self-consistently. 
This bond field $\Delta$ naturally appears in our path integral formalism \cite{bejas12} as 
a Hubbard-Stratonovich field. $N_s$ is the total number of lattice sites and 
$f(x)=1/({\rm e}^{x/T} +1)$ is the Fermi-Dirac distribution function. 
The electron dispersion $\epsilon_{\vk}$ is renormalized already at leading order in the large-$N$ 
expansion 
\be
\epsilon_{\vk}= -2 \left( t \frac{\delta}{2} + J \Delta \right) (\cos k_x+\cos k_y)-
4 t'\frac{\delta}{2}\cos k_x \cos k_y - \mu,
\ee
where $t$ and $t'$ are reduced by a factor of $\delta /2$ and $\mu$ is the chemical potential. 
The $d$-wave polarization $\Pi_{d}(\vq,\omega)$ in \eq{chid} reads 
\be
\Pi_{d}(\vq, \omega) = - \frac{1}{N_{s}}\;
\sum_{\vk}\; \gamma^2(\vk) \frac{f(\epsilon_{\vk + \vq/2}) 
- f(\epsilon_{\vk - \vq/2})} 
{\epsilon_{\vk + \vq/2} - \epsilon_{\vk- \vq/2}-\omega - {\rm i}\Gamma}\,,
\label{Pid}
\ee
where the $d$-wave form factor $\gamma(\vk)=(\cos k_x - \cos k_y)/2$ describes a $d$-wave symmetry 
of the bond-charge order and $\Gamma$ is infinitesimally small. 
In the limit of $\vq={\bf 0}$, $\chi_d(\vq,\omega)$ would be reduced to the 
electronic nematic susceptibility \cite{yamase04b} associated with a $d$-wave Pomeranchuk instability \cite{yamase00a,yamase00b,metzner00}. 
In the following, we measure all quantities with the dimension of energy in units of $t$. 
A realistic value of $t/2$ (Ref.~\onlinecite{misc-t}) in cuprates is 
around 500 meV (Ref.~\onlinecite{hybertsen90}),

\section{Results} 

\begin{figure} [th]
\centering
\includegraphics[width=8cm]{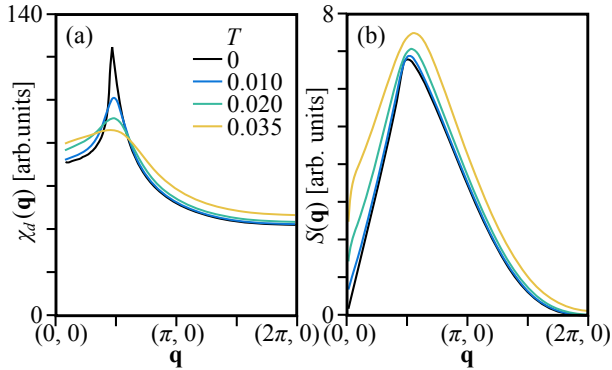}
\caption{
(Color online) 
(a) $\vq$ dependence of $\chi_d(\vq)$ for various choices of temperatures at $\delta=0.13$. 
(b) $S(\vq)$ for the same parameters as in (a) and the cutoff energy is $\omega_c=\infty$.  
}
\label{Gamma0001}
\end{figure}

We choose $J=0.3$ and $t'=0.3$ in our Hamiltonian (\ref{tJV}) as typical parameters 
for e-cuprates \cite{bejas14}; the precise value of $V$ is not important as long as it suppresses 
phase separation. As a value of $\Gamma$ in \eq{Pid} 
we take $\Gamma=10^{-4}$, which is reasonably small. 
In \fig{Gamma0001}(a), we present the static part of the $d$-wave bond-charge 
susceptibility $\chi_{d}(\vq)=\chi_{d}(\vq,\omega=0)$ as a function of $\vq$ for several choices 
of temperatures at $\delta=0.13$. 
Note that $\chi_{d}(\vq)$ has $4\pi$ periodicity because of the presence of the $d$-wave form factor 
[see \eq{Pid}] and thus the $\vq$ region is in $0 \leq q_x \leq 2\pi$ in  \fig{Gamma0001}(a). 
With decreasing $T$, a peak is pronounced at $\vq=(\pm 2 \pi Q_{\rm co}, 0)$ and 
$(0,\pm 2 \pi Q_{\rm co})$ with $Q_{\rm co}\approx 0.25$, indicating a tendency 
toward a charge ordered phase. However, the static susceptibility does not diverge 
and the CO remains a short range. 

We show in \fig{Gamma0001}(b) 
the equal-time correlation function $S(\vq)$ for the same parameters as in \fig{Gamma0001}(a). 
While $S(\vq)$ has a peak at almost the same position of $\chi_{d}(\vq)$, the peak intensity 
is slightly {\it suppressed} with decreasing $T$ (Ref.~\onlinecite{misc-EPL}) 
even though $\chi_{d}(\vq)$ shows a pronounced peak at low $T$ [\fig{Gamma0001}(a)]. 
This counter-intuitive feature comes from the presence of the Bose factor $n_{B}(\omega)$ in \eq{Sq}.  
In fact, if $n_{B}(\omega)$ were omitted in \eq{Sq}, $S(\vq)$ would show a peak, which 
is enhanced with decreasing $T$.  

The suppression of $S(\vq)$ with decreasing $T$ in \fig{Gamma0001}(b) can be an artifact 
because we completely neglect the damping of quasiparticles, that is, we assume $\Gamma=10^{-4}$ 
at any temperature. Apparently this is unphysical. 
In fact, a relatively large $\Gamma$ is frequently assumed when a spectral line shape 
obtained theoretically is compared with experimental data \cite{greco16,ishii17,greco18}. 
For example, in Ref.~\onlinecite{greco16} a comparison 
to the experimental data was made successfully by assuming that $\Gamma$ increases 
with increasing temperature to understand 
the temperature dependence of the high-energy charge-excitation spectrum.  
Similarly, the damping effect should be pronounced also in a low doping region 
because of strong antiferromagnetic fluctuations there in cuprates. Therefore, 
we invoke a finite value of $\Gamma$ in \eq{Pid}
to simulate phenomenologically the damping of quasiparticles 
as a broadening of the spectrum.  

\begin{figure} [th]
\centering
\includegraphics[width=8cm]{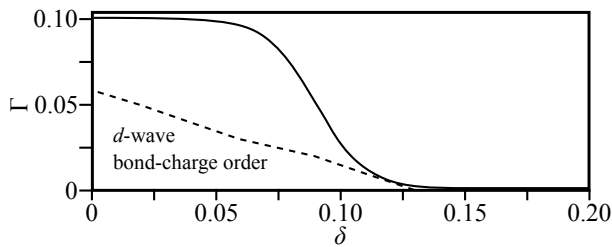}
\caption{
Doping dependence of the damping $\Gamma$  (solid line) and 
the phase boundary of the $d$-wave bond-charge order at $T=0$ (dashed line).  
}
\label{Gamma-delta}
\end{figure}

A value of $\Gamma$ may in principle depend on $T$, $\delta$, $\vq$, $\omega$, and others. 
Since our major interest is a study of temperature and doping dependences of $S(\vq)$, 
we consider only possible $T$ and $\delta$ dependences of $\Gamma$. 
As a function of $T$, a leading correction may be given by a linear term in $T$ (Ref.~\onlinecite{misc-MFL}), 
i.e., 
\be
\Gamma(\delta,T) = \Gamma(\delta) + \alpha T \,.
\label{Gamma-T}
\ee
Concerning the doping dependence, we recall that 
neutron scattering experiments \cite{motoyama07} revealed that the antiferromagnetic 
correlation length starts to increase substantially below $\delta \approx 0.10$ in the normal 
metallic phase around $T \sim 300$ K. 
Concomitantly, quasiparticles may be damped heavily below $\delta \approx 0.10$. 
To mimic this phenomenology in a simple way, we assume 
a $\delta$ dependence of 
$\Gamma(\delta)$ as shown in \fig{Gamma-delta}, where $\Gamma$ increases rapidly 
below $\delta \approx 0.10$. An explicit expression is given by 
\be
\Gamma(\delta) =0.001 + 0.05\left[ 1- \tanh \left(\frac{\delta-0.09}{0.02}\right) \right].
\label{Gamma-form}
\ee

In \fig{Gamma-delta} we also plot the boundary of the $d$-wave bond-charge phase 
at $T=0$. When $\Gamma$ is infinitesimally small and independent of doping, the model would 
exhibit the $d$-wave bond-charge 
instability at $\delta_c \approx 0.125$ (Ref.~\onlinecite{bejas14} and also see Appendix). 
With increasing $\Gamma$, 
the $d$-wave bond-charge phase shrinks. 
As a result, we have only charge fluctuations associated with the $d$-wave 
bond-charge order for doping above the dashed line in \fig{Gamma-delta}. 

While the choice of the absolute value of $\Gamma$ is rather arbitrary in \eq{Gamma-form}, 
we choose it to have no charge instabilities even at low doping rate at $T=0$ (solid line in 
\fig{Gamma-delta}), so that 
our calculations are performed in the paramagnetic phase in the entire doping region. 
For a finite $T$, we choose $\alpha=9$ in \eq{Gamma-T} after checking that 
our conclusions are not modified for other choices of $\alpha=3$ and $6$. 
While we specified the functional form of $\Gamma$ [Eqs.~(\ref{Gamma-T}) 
and (\ref{Gamma-form})] to perform systematic  calculations, 
the precise functional form itself is not important as long as $\Gamma$ increases 
with increasing temperature and decreasing  doping, so that charge 
instabilities (see Appendix for details) are suppressed in a low doping region.

\begin{figure}[htb]
\centering
\includegraphics[width=7cm]{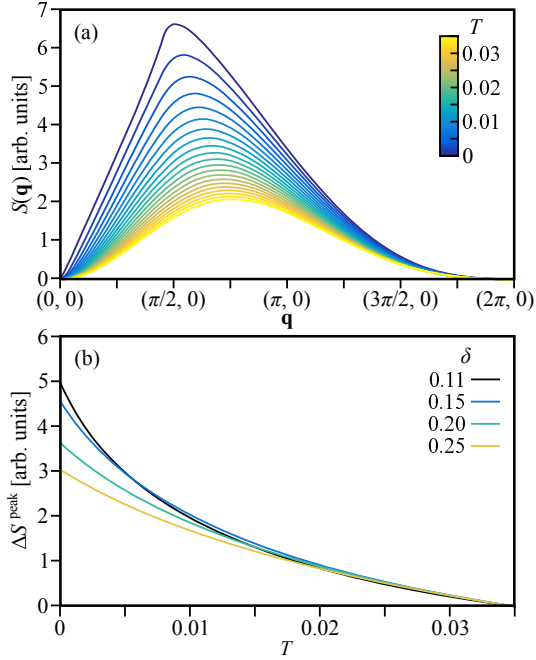}
\caption{(Color online) 
(a)  $\vq$ dependence of $S(\vq)$ 
for various choices of temperatures at doping $\delta=0.13$. 
(b) $\Delta S^{\rm peak}$ as a function of temperature for several choices of doping rates.  
The cutoff energy is $\omega_c=\infty$. 
}
\label{q-Sq}
\end{figure}

Figure~\ref{q-Sq}(a) shows $S(\vq)$ along the direction $(0,0)$-$(2\pi,0)$ 
in a temperature range $0<T<0.035$ for $\delta=0.13$. 
$S(\vq)$ forms a peak structure 
at $\vq \sim (0.5\pi,0)$ even at high $T$. 
To clarify how the peak of $S(\vq)$ develops, 
we define $\Delta S(\vq) = S(\vq; T) - S(\vq; T=0.035)$. 
Since $T=0.035$ corresponds to $T\approx 400$ K, 
the intensity at $T=0.035$ may be regarded as a {\it background}. 
Hence $\Delta S(\vq)$ is regarded as $S(\vq)$ after background subtraction, as often seen 
in an experimental data analysis \cite{da-silva-neto16}.  
We plot in \fig{q-Sq}(b) the temperature dependence of the peak intensity of 
$\Delta S(\vq)$ for several choices of doping rates ($>0.10$). 
The peak intensity $\Delta S^{\rm peak}$ is pronounced upon decreasing temperature and 
doping rate, which is qualitatively the same as the experimental observation 
(see Fig. 2~G in Ref.~\onlinecite{da-silva-neto16}).

\begin{figure}[htb]
\centering
\includegraphics[width=8cm]{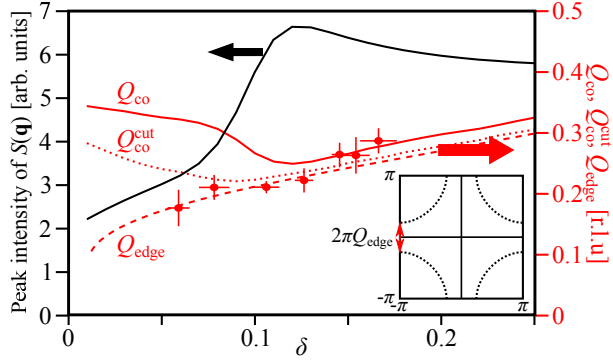}
\caption{(Color online) 
Peak intensity of $S(\vq)$ with $\omega_c=\infty$ 
and the momenta $Q_{\rm co}$, $Q_{\rm co}^{\rm cut}$, and  $Q_{\rm edge}$ 
as a function of $\delta$ at $T=0$. $Q_{\rm co}$ is the peak position of 
$S(\vq)$ with $\omega_c=\infty$ whereas $Q_{\rm co}^{\rm cut}$ is that of 
$S(\vq)$ with $\omega_c=0.05$; $Q_{\rm edge}$ is defined in the inset. 
Solid circles are the experimental data from Ref.~\onlinecite{da-silva-neto16}.  
}
\label{delta-Sq}
\end{figure}

With decreasing $\delta$ beyond $\delta\approx 0.10$, $S(\vq)$ is suppressed substantially 
as shown in \fig{delta-Sq}, which is actually observed in experiments \cite{da-silva-neto16}. 
This rapid suppression comes from the pronounced increase 
of the damping $\Gamma$ below $\delta \approx 0.10$ as shown in \fig{Gamma-delta}. 
If $\Gamma$ is assumed to be constant, 
the peak intensity of $S(\vq)$ would continue to increase 
with decreasing $\delta$. 

As shown in Fig.~\ref{q-Sq}(a), 
$S(\vq)$ exhibits a peak at $\vq=(\pm 2\pi Q_{\rm co}, 0)$ and $(0, \pm 2\pi Q_{\rm co})$. 
This peak position is plotted in \fig{delta-Sq} as a function of doping rate at $T=0$ together with 
$Q_{\rm edge}$, the distance between the Fermi surface edges across $\vk=(\pi,0)$ and 
its equivalent wavevectors (see the inset in \fig{delta-Sq}). 
As pointed out in Ref.~\onlinecite{yamase15b}, the peak structure is formed 
by particle-hole scattering processes characterized by $Q_{\rm edge}$. 
Hence $Q_{\rm co}$ corresponds to 
such scattering wavevector $Q_{\rm edge}$ at least down to $\delta\approx 0.10$, 
although it becomes slightly larger than $Q_{\rm edge}$, 
since $S(\vq)$ is an energy-integrated quantity [see \eq{Sq}].  
Below $\delta \approx 0.10$, $Q_{\rm co}$ goes up and deviates substantially from 
$Q_{\rm edge}$. This is because the damping $\Gamma$ increases rapidly and the 
structure coming from the underlying Fermi surface is blurred.

\begin{figure}[htb]
\centering
\includegraphics[width=8cm]{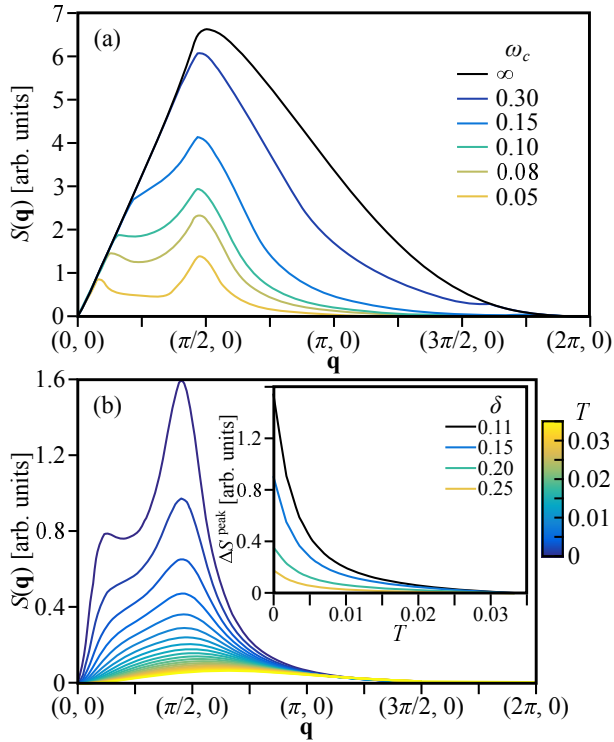}
\caption{(Color online) 
(a) $\vq$ dependence of $S(\vq)$ for several choices of the cutoff energy at $\delta=0.13$ and $T=0$.
(b) $\vq$ dependence of $S(\vq)$ with a cutoff energy $\omega_c=0.05$ for 
various choices of temperatures at $\delta=0.13$. 
The inset shows the temperature dependence of the peak intensity of $\Delta S$ 
for several choices of doping rate after subtraction of intensity at $T=0.035$. 
}
\label{Sq_cut}
\end{figure}

The peak width of $S(\vq)$ in \fig{q-Sq}(a) is very large even at $T=0$. 
This broadness is not due to our introduction of a finite $\Gamma$ [Eqs.~(\ref{Gamma-T}) and 
(\ref{Gamma-form})].  In fact, the peak of $S(\vq)$ is very broad even 
for $\Gamma=10^{-4}$ as shown in \fig{Gamma0001}(b), 
although the static susceptibility exhibits a sharp peak at $\vq=(0.5\pi,0)$ at low $T$. 
This counter-intuitive feature originates from the $\omega$ integration in \eq{Sq}. 
To demonstrate this, we decrease the cutoff energy $\omega_c$ in \eq{Sq}. 
As shown in \fig{Sq_cut}(a), the resulting spectrum exhibits a shaper peak 
around $\vq=(0.5\pi,0)$ for lower $\omega_c$. 
That is, the sharp peak of $S(\vq)$ for low $\omega_c$ originates from 
the short-range $d$-wave bond-charge order and this sharp peak is simply smeared out 
by spectral weight coming from the high-energy region. 
To extract direct contributions from the short-range $d$-wave bond-charge order, therefore, 
it may make sense to consider $S(\vq)$ with a low cutoff energy $\omega_c$, 
as actually done in a recent experimental analysis \cite{da-silva-neto18}. 
In \fig{Sq_cut}(b), we plot $S(\vq)$ for $\omega_c = 0.05$ for various choices of temperatures. 
$S(\vq)$ exhibits a broad spectrum 
at high $T$, but a sharp peak gradually grows below $T \sim 0.02$ as a consequence of 
development of a short-range bond-charge order. 
Similar to \fig{q-Sq}(b), we plot $\Delta S^{\rm peak}$ as a function of $T$ for several choices of doping 
in the inset of \fig{Sq_cut}(b). 
The overall feature is the same as \fig{q-Sq}(b), that is, the peak intensity 
increases at lower $T$ and for lower doping, consistent with the experiments \cite{da-silva-neto16}. 
Denoting the peak position of $S(\vq)$ in \fig{Sq_cut}(b) by $Q_{\rm co}^{\rm cut}$, 
we plot its doping dependence in \fig{delta-Sq}. 
$Q_{\rm co}^{\rm cut}$ tends to follow $Q_{\rm edge}$ down to rather low doping and 
starts to go upward below $\delta \approx 0.1$.

\section{Discussions}
The equal-time correlation function $S(\vq)$ can be measured directly by RXS. 
In particular, we computed $S(\vq)$ associated with the $d$-wave bond-charge order, 
which is reduced to the nematic order in the limit of $\vq={\bf 0}$ (Refs.~\onlinecite{yamase00a,yamase00b,metzner00}). 
A possibility of $d$-wave bond-charge order \cite{comin15}  
as well as the nematic order \cite{hinkov08,daou10,cyr-choiniere15,sato17} 
is already discussed in h-cuprates. 
Therefore it is a crucial step toward the understanding of the charge dynamics in cuprates 
to clarify to what extent 
our results capture recent RXS data in e-cuprates \cite{da-silva-neto15,da-silva-neto16,da-silva-neto18}. 
Since the previous data of Ref.~\onlinecite{da-silva-neto15} were already 
discussed in Ref.~\onlinecite{yamase15b}, 
we focus on the recent experimental data reported in Refs.~\onlinecite{da-silva-neto16} 
and  \onlinecite{da-silva-neto18}.

The peak width of $S(\vq)$ shown in \fig{q-Sq}(a) is much larger than the experimental one 
\cite{da-silva-neto16}. 
However, when we focus on a low-energy region as shown in \fig{Sq_cut}(b), 
the peak width becomes substantially small especially below $T \sim 0.01 (\approx 100$ K)   
and comparable to the experimental data \cite{da-silva-neto16}. 
This suggests that compared with experiments, our results in \fig{q-Sq}(a) emphasize too much 
the contributions from 
high-energy spectral weight in $S(\vq)$ [\eq{Sq}]. 
We consider two reasons for that. 
First, the spectral weight in a high-energy region in experiments 
can be much weaker than the present results  
and thus our results overestimates such a contribution, leading to a much broader peak. 
Second, as found in Ref.~\onlinecite{da-silva-neto18}, 
the spectral weight reported in Refs.~\onlinecite{da-silva-neto15} and \onlinecite{da-silva-neto16} 
comes not only from charge excitations, but also from magnetic excitations. In particular, 
the magnetic contribution extends from 200 to 700 meV at $Q_{\rm co}$ and 
is pronounced below $T \sim 300$ K.  
These contributions are surely important to the resulting line shape of the spectrum, but 
are fully neglected in the present theory. 

Nevertheless, our results are in line with the recent RIXS measurements. 
In Ref.~\onlinecite{da-silva-neto18}, the spectral weight related to the CO 
signal is discussed to come from energy below 60 meV, 
whereas our sharp peak is realized in \fig{Sq_cut}(a) for $\omega_c \approx 0.05 - 0.1$, 
which corresponds to 50 - 100 meV. 
A possible reason why the spectral weight of charge excitations 
concentrates only in the low-energy region in experiments is that the actual system is located much 
closer to the vicinity of CO phase than the present theoretical situation 
(see \fig{Gamma-delta}). 

The experimental observation that the energy range of charge dynamics 
coincides with that of magnetic excitations \cite{da-silva-neto18} 
is also in line with our theory based on $d$-wave bond-charge order. 
As shown in Refs.~\onlinecite{bejas12} and \onlinecite{bejas17}, 
$d$-wave bond-charge order comes from 
the exchange interaction, i.e., the $J$-term in the $t$-$J$ model. 
This is also clear from \eq{chid} where the interaction strength $J$ enters. 
If the charge dynamics originates from usual on-site charge excitations, 
its major contribution would appear in a higher energy region as 
plasmons \cite{greco16,bejas17,greco18}.  On the other hand, needless to say, magnetic excitations are 
controlled by the $J$-term. Therefore both bond-charge and magnetic excitations naturally appear 
in the same energy scale of $J$. Obviously more theoretical studies are necessary 
when one wishes to address more details: for example, the reason why the magnetic excitations are 
strongest around $\omega = 0.2$ eV whereas the typical charge excitation energy is lower 
than that, and why the magnetic excitations are enhanced at the {\it same} wavevector as that of the 
bond-charge excitations \cite{da-silva-neto18}. 
These issues are beyond the scope of the present work.

The peak position of $Q_{\rm co}$ tends to be larger with increasing $T$ as shown in \fig{q-Sq}(a) 
whereas the peak position is almost independent of $T$ in the experiments \cite{da-silva-neto16}. 
However, given that the magnetic excitations are included in $S(\vq)$ mainly in a high-energy part 
in the experiments  \cite{da-silva-neto16}, a comparison to the experiments may be made 
better by focusing on a low-energy region as shown in \fig{Sq_cut}(b). 
In fact, in this case, our peak position is almost independent of $T$ 
below $T\sim 0.01$, which corresponds to about 100 K.

$Q_{\rm co}$ decreases with decreasing doping in the experiments \cite{da-silva-neto16} 
(the experimental data are plotted in \fig{delta-Sq}). 
In particular, it tends to be constant in $\delta \lesssim 0.10$ if we do not consider seriously 
the data at $\delta=0.059$, where the existence of a peak at $Q_{\rm co}$ in Fig. 2(B) 
of Ref.~\onlinecite{da-silva-neto16} is unclear.  
We think that our theoretical results shown in \fig{delta-Sq} reasonably capture   
the major feature of $Q_{\rm co}$ 
observed in the experiments. In particular, if we consider  $Q_{\rm co}^{\rm cut}$, instead of $Q_{\rm co}$, 
the agreement with the experimental data is more satisfactory in a wide doping region.

As shown in \fig{delta-Sq}, the peak intensity of $S(\vq)$ gradually increases 
and sharply drops in $\delta \lesssim 0.10$. Therefore the CO signal 
is expected to become difficult to be detected in $\delta \lesssim 0.10$.  Conversely, we expect that 
the CO signal is observed  more clearly above $\delta \approx 0.10$, as is indeed the 
case in the experiments \cite{da-silva-neto16}. 

In the region above $\delta \approx 0.10$, the peak intensity of $S(\vq)$ increases at lower temperature 
and decreases for higher doping as shown in the inset of \fig{Sq_cut}(b). This tendency 
is the same even if we take $\omega_c=\infty$ and include high-energy contribution to $S(\vq)$ 
[\fig{q-Sq}(b)]. Those temperature and doping dependence are consistent with the 
experimental observation \cite{da-silva-neto16}.  
Although the peak intensity in \fig{Sq_cut}(b) seems to develop below a certain temperature, 
it should not be interpreted as the onset of the $d$-wave bond-charge order 
because the static susceptibility does not diverge down to zero temperature [see \fig{Gamma0001}(a)].

We have shown that $d$-wave bond-charge order can reasonably capture the major features of 
low-energy charge excitations in e-cuprates \cite{da-silva-neto15,da-silva-neto16,da-silva-neto18}. 
What about other types of CO? The crucial points in the experimental data are twofold: 
(a) the CO tendency is detected along the $(0,0)$-$(\pi,0)$ direction and 
(b) its momentum is controlled mainly by $2 k_F$ scattering processes across the Fermi surface 
edges along the Brillouin zone boundary (inset of \fig{delta-Sq}). 
As shown in Figs.~4, 6, and 7 in Ref.~\onlinecite{bejas17}, 
the first feature (a) cannot be captured by other promising types of CO obtained in 
the $t$-$J$ model such as $s$-wave bond-charge order 
and flux phase whereas the $d$-wave bond-charge order can capture it.  
There are various $2 k_F$ scattering processes for the Fermi surface shown 
in the inset of \fig{delta-Sq}. 
The reason why the scattering processed near $(\pi,0)$ and $(0,\pi)$ is favored lies 
in the $d$-wave form factor, which is enhanced around $(\pi,0)$ and $(0,\pi)$. 
In this sense, the feature (b) is in line with the presence of a $d$-wave form factor.

\section{Summary} 
Motivated by the recent RXS measurements of the doping dependence 
of the short-range CO observed in e-cuprates \cite{da-silva-neto18}, 
we studied  the equal-time correlation function $S(\vq)$ associated with 
the $d$-wave bond-charge order by using a large-$N$ technique in the two-dimensional 
$t$-$J$ model.
We extended our previous work \cite{yamase15b}  
for a single doping rate by introducing the doping and temperature dependences of 
the damping $\Gamma$ and showed 
the consistency of our proposed scenario of the $d$-wave bond-charge order. 
The short-range CO is pronounced at low temperature (Figs.~\ref{q-Sq} and \ref{Sq_cut}).  
The peak intensity develops gradually down to $\delta \approx 0.10$ and is substantially 
suppressed below that (\fig{delta-Sq}), due to the strong damping effect presumably 
coming from antiferromagnetic fluctuations. 
A recent experiment on a h-cuprate \cite{jang18} also shows 
that the CO competes with antiferromagnetic fluctuations.
The momentum of the short-range CO decreases with lowering doping and goes up 
below $\delta \approx 0.10$ (\fig{delta-Sq}). All these features reasonably capture the 
essential features of the recent experimental data \cite{da-silva-neto16}. 
This agreement suggests three important implications for the physics in e-cuprates. 
(i) The origin of the CO lies in the magnetic exchange interaction, 
i.e., $J$-term in the $t$-$J$ model.  
(ii) The CO is not generated by antiferromagnetic fluctuations. Rather they seem to 
contribute to the enhancement of the quasiparticle damping and consequently to 
suppress the charge ordering tendency. 
(iii) The CO is not a usual charge-density-wave, but a bond-charge order. In particular,  
it is characterized by $d$-wave symmetry and is connected to the nematic order in the 
limit of $\vq={\bf 0}$. In this sense, the nematic physics plays a role also 
in e-cuprates, although so far it has been discussed only in h-cuprates.

\acknowledgments
The authors thank E. H. da Silva Neto, K. Ishii, B. Keimer, M. Minola, T. Tohyama, and R. Zeyher 
for very fruitful discussions.
H.Y. acknowledges support by JSPS KAKENHI Grant Number JP15K05189. 
A.G. acknowledges the Japan Society for the Promotion of Science
for a Short-term Invitational Fellowship program (S17027) under which this work was initiated.

%\newpage

\appendix
\section{$\boldsymbol{S(\vq)}$ along the diagonal direction}
\begin{figure} [th]
\centering
\includegraphics[width=7cm]{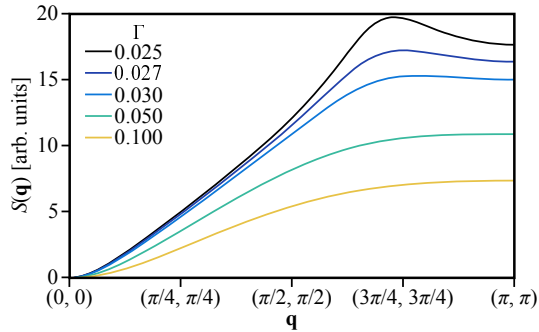}
\caption{$S(\vq)$ along the $(0,0)$-$(\pi,\pi)$ direction for various choices of 
$\Gamma$ at $T=0$ and $\delta=0.08$; the cutoff energy is $\omega_c = \infty$.  
}
\label{Sq-diagonal}
\end{figure}

The $d$-wave bond-charge order in \fig{Gamma-delta} occurs at  
$\vq_{1} = (\pm q_1,  \pm q_1)/\sqrt{2}$ and 
the instability at $\vq_{2}= (\pm 2 \pi Q_{\rm co}, 0)$ and $(0,\pm 2 \pi Q_{\rm co})$ is 
the second leading one. Nevertheless, the peak 
structure of $S(\vq)$ as well as $\chi_{d}(\vq)$ becomes sharper at $\vq_{2}$ with decreasing $T$ whereas 
the spectrum around $\vq_{1}$ is typically very broad and the peak structure develops only in the vicinity 
of the onset temperature of the charge instability. This peculiar feature was addressed in detail 
in Ref.~\onlinecite{yamase15b}. 
In the present paper, we introduced a large $\Gamma$. In this case, 
the peak structure around $\vq_{1}$ is not realized unless the system is located in the 
vicinity of the phase boundary of the $d$-wave bond-charge order (dashed line in \fig{Gamma-delta}). 
To demonstrate this, we compute $S(\vq)$ for various choices of $\Gamma$ along the 
$(0,0)$-$(\pi,\pi)$ direction for $\delta=0.08$ at $T=0$ in \fig{Sq-diagonal}. 
There is a broad structure for a large $\Gamma$ 
and a small peak develops around $(0.75\pi, 0.75\pi)$ 
only near $\Gamma=0.025$, which is very close to the phase boundary. 
Since a large momentum near $\vq \approx (\pi,\pi)$ is not accessible by RXS and 
furthermore the peak structure around $\vq_{1}$ is not realized in general in the presence of a large $\Gamma$, 
we focus in this paper on the peak structure around $\vq_{2}$, which is relevant to RXS as well as to recent experimental 
data \cite{da-silva-neto16,da-silva-neto18}. 

\bibliography{main} 

\begin{thebibliography}{59}
\bibitem[{mis({\natexlab{a}})}]{misc-d-wave-bond}
\bibinfo{note}{The $d$-wave bond-charge order is referred to as various terms
  in literature, depending on its modulation vector $\vq$: kite state when
  $\vq$ is close to $(\pi,\pi)$ in Refs.~\onlinecite{affleck88a} and
  \onlinecite{cappelluti99}, $d$-wave Pomeranchuk instability when $\vq$ is
  close to $(0,0)$ in Refs.~\onlinecite{bejas12}, \onlinecite{yamase00a},
  \onlinecite{yamase00b}, and \onlinecite{metzner00}, and BOP$_{x \bar{y}}$ for
  a general $\vq$ in Ref.~\onlinecite{bejas14}. The $d$-wave bond-charge order
  does not break time-reversal symmetry and thus should not be mixed with flux
  phase \cite{affleck88a,cappelluti99}, i.e., the so-called $d$-density wave
  \cite{chakravarty01} or $d$-wave charge density wave \cite{bejas12}, which
  breaks time-reversal symmetry. In the present paper we adopt the term of
  $d$-wave bond-charge order to emphasize the $d$-wave symmetry and bond
  character.}
\bibitem[{mis({\natexlab{b}})}]{misc-t}
\bibinfo{note}{A factor of $1/2$ comes from a large-$N$ formalism where $t$ is
  scaled by $t/N$ and the physical situation corresponds to $N=2$.}
\bibitem[{mis({\natexlab{c}})}]{misc-EPL}
\bibinfo{note}{In Fig.~3(d) in Ref. ~\onlinecite{yamase15b} this feature is
  less pronounced because the temperature dependence of the numerator in
  \eq{chid} was not taken into account.}
\bibitem[{mis({\natexlab{d}})}]{misc-MFL}
\bibinfo{note}{Stimulated by the marginal Fermi liquid theory \cite{varma96},
  one might wonder how our results are modified if $\Gamma$ depends also on
  $\omega$ by assuming $\Gamma \propto {\rm max}(T, \omega)$ at fixed density.
  In this case, the peak intensity and the peak position of $S(\vq)$ may
  change, but our conclusions are not modified.}
\end{thebibliography}

\end{document}